\documentstyle[12pt,epsfig]{article}
\begin{document}
\begin{center}
\vspace{1.5in}
{\LARGE Colour Confinement, Baryon Asymmetry and Dark Matter in 
Quantum Chromodynamics }
\end{center}
\vspace{.4in}
\begin{center}
{\bf Afsar Abbas}\\
Institute of Physics\\ 
Bhubaneshwar-751005, India\\
\vspace{,1in}
email: afsar@iopb.res.in
\end{center}
\vspace{.5in}
\begin{center}
{\bf Abstract}
\end{center}
\vspace{.3in}

The concepts of local and global colour singletness in QCD 
are analyzed within a framework of a group theoretical technique  
which can project out various colour representations of the QGP. 
A proof of colour singletness is thereby provided. This also leads to 
an explanation of the baryon asymmetry in the universe within the 
framework of QCD itself. Supersymmetry plays a basic role in this 
explanation as well as providing us with a consistent connection between 
the current and the constituent picture of quarks. Glueballs as a 
Dark Matter candidate manifest themselves herein. Baryonlessness of QGP 
is a clear prediction of this model.

\newpage

Colour confinement is an outstanding problem of quantum chromodynamics.
In spite of intense efforts one still does not know as to how it
arises in QCD and even comparisons with the intrinsic difficulties of 
the  Fermat's Last Problem are being made [1, p.291]. 
The equally significant problem of baryon asymmetry in the
universe is generally believed not to arise from QCD. One invokes exotic
frameworks to understand this phenomena. In
this paper however, it shall be shown that actually the colour confinement
and baryon asymmetry are correlated and can be understood within QCD.
This will help explain the current and constituent quark connection.
Supersymmetry shall appear as basic to this understanding. This will also 
help in solving the Dark Matter conundrum.

It is an experimental fact that as of now no convincing case of free
coloured objects has yet been discovered. This is the primary reason for
invoking the colour confinement hypothesis in QCD. Though no violation of
the same has yet been found, the idea is still referred to as a 
"hypothesis" as the same has not been demonstrated to follow from QCD. 
This is essentially a 'local' colour confinement hypothesis. 'Local' is to 
sense that all known hadrons of size less than 
$\sim 1-2 $ fm confine quarks and gluons 
and are believed to be in the colour singlet state.
       
However with the advent of the possibility of quark-gluon plasma it
is believed that at high enough temperatures and densities the quarks and
gluons become deconfined. Deconfined, with respect to the 'local' colour
confinement concept stated above. It is commonly believed that this QGP
should still be colour confined 'globally'. In fact, QGP is understood to
be a transition from a local colour singlet framework to a global colour
singlet framework [2]. In the theory of Heavy Ion Collision scenario of 
the creation of QGP the 'global' prefix really means that at the 
scale  of two heavy colliding and overlapping nuclei in sizes of 
$\sim 10-20$ fm the QGP is globally confined.

If there be 'quark stars' of size of a few kilometres, as in some
astrophysical scenarios, then this too is expected to be globally colour
singlet over this large size [3]. In fact this concept of global colour
singletness is routinely extrapolated to invoke colour singletness
for the whole universe as well!
       
The empirical evidence for confinement as stated above is lack of any 
free coloured object in various searches. 
However. this can easily be understood as a result of 
the 'local' colour singlet hypothesis alone. That is  
just to explain this one need not invoke global colour singlet 
hypothesis. Note that there are no astrophysical scenarios where QGP is 
being generated spontaneously ( though we are trying to create QGP in the 
laboratory artificially ) today. Also there are no known
clear cut cases of quark stars. Therefor there is no clearcut empirical
support for the 'global' colour hypothesis. In fact if a quark matter were
to exist at the centre of a neutron star then one need not invoke any 
colour confinement hypothesis as these quarks would then be confined 
automatically by the pressure of the surrounding matter.
The situation is similar to the "little bag" model of hadrons ( popular  
sometime ago ) where the quarks were confined inside baryons by the 
pressure of the pions outside. So given all the astrophysical realities, 
the "global" colour hypothesis should basically be treated as an 
assumption. As such it may be put to test empirically.

Hence to distinguish these confinement scenarios it would be appropriate 
to classify as "strong" confinement hypothesis as the one which
just incorporates the 'local' colour hypothesis. But at large distances 
it does not necessarily invoke global colour hypothesis. So in principle 
the 'strong' confinement hypothesis takes the philosophy that we really do 
not know as to whether colour singletness is guaranteed globally or not 
and hence we allow for other colour representations to play a role if 
necessary. The prefix 'strong' here is used to take account of the fact 
that as far as we know for sure today ( ie. empirically ) local colour 
singleteness is all that we are sure of.   
In the same spirit let us label the hypothesis which incorporates global
colour singlet hypothesis as 'weak' confinement hypothesis. The
prefix 'weak' to account for the fact that no specific empirical
evidence of the same exists and that it is actually a theoretical 
assumption. Note though that in the limit of size
1-2 fm the "global" colour confinement hypothesis would collapse to the 
"local" colour confinement hypothesis.
        
Let us next ask whether we can completely ignore the 'weak' 
confinement hypothesis and just use the 'strong' confinement
hypothesis. What this would mean is that globally the QGP (like the one we
are trying to create in heavy ion collisions in the laboratory at present)  
may actually be globally colour non-singlet.
Coloured objects would not leak out of this
global coloured QGP due to the fact that all the particles would be
confined due to the dynamics of the collision. Later as the QGP cools it
will produce only locally colour singlet objects. By demanding the 'weak' 
confinement hypothesis one may be invoking a spurious global colour 
confinement and thereby be missing some essential physics. As we shall show 
below, indeed we are missing quite a lot of physical reality by doing so!

To understand this let us consider a bunch of quarks, anti-quarks and
gluons at some finite temperature T. We can obtain partition function for
the same and thereby obtain various quantities of interest in QGP [2].
One uses the powerful mathematical development of the consistent inclusion 
of the internal symmetries ( like  $S U (3)_{c}$ of QCD ) 
in a statistical thermodynamic description of the quantum gas.
This group theoretical technique ( described in Appendix below ) has been 
used by other groups to project out colour singlet representations of the 
QGP. This projection technique necessarily projects out the global
colour singlet states of QGP [2,4,5]. Thereby the significance of global
colour singleteness condition on QGP was pointed out by the authors  
[2,4,5,6]. They studied the ratio of the colour singlet projected to the 
non-projected energy of the QGP. This is plotted as a function of $T 
{V}^{1/3}$ (see Figure 1 for this plus other aspects to be discussed 
below). At high  $T {V}^{1/3}$ values the two states, the
colour singlet projected and the non-projected, are degenerate.
At smaller $T {V}^{1/3}$ there is a drop off
( as in a phase transition ) of the colour singlet states.
The physical interpretations are given elsewhere [2,4,5,6]. Here our main
intention is to study the degeneracy at high  $T {V}^{1/3}$ 
and the sudden drop off of the singlet states at small  $T {V}^{1/3}$.

All this was for the "global" colour singlet states.
A couple of years ago we studied this problem in detail.
Suppose globally the QGP could be in some coloured state?
How will this manifest itself? The same group
theoretical technique ( see Appendix ) was powerful enough to impose 
"global" colour representations like the 8-plet, the 27-plet etc.
These were studied by us sometime ago [7] and for the sake of 
completeness and clarity a similar plot is reproduced here as Figure 1.
Again the ratio of different colour states is plotted. 
The result and discussion has already been presented by us earlier [7].
Here we would like to re-analyze the information in the light of
the 'strong' and the 'weak' confinement conditions. This will allow us 
to extract further deep physics not perceived earlier [7]. 

Note that all this was done for total baryon number zero. Others have 
tried to incorporate additional explicit baryon number [8,9] to project 
colour singlet states. They could do so only by extending the group to 
$U(1)_{A} \bigotimes S U (3)_{c}$ and that too
for the colour unprojected and the colour singlet states only.
The point to note is that inclusion of an 
explicit baryon excess necessitates the enlargement of the 
QCD group. But this is actually going beyond QCD. 
As we wish to stick to the QCD $S U (3)_{c}$ group 
we find that there is no way that one can have baryon excess in
the QGP of $S U (3)_{c}$

Note from Fig.1 that all the lines converge and merge at high 
$ TV^{1/3} $. This means that for large values of $TV^{1/3}$
all representations: singlet, octet, 27-plet etc are degenerate with the
unprojected energy ( here the 27-plet state is not shown but has similar 
behaviour ). For sufficiently large
QGP and sufficiently high temperatures all the energies of all
the representations are found to be
degenerate. Globally there is nothing which favours the colour-singlet 
representation over the colour-octet at high temperatures. 
Note that this result could directly be seen from 
the expression $Z_{(p,q)}$ which 
for the continuum approximation for sufficiently
large volume and temperature becomes independent of representation [7].
Hence energies for each colour representation are degenerate with 
each other for large $ TV^{1/3} $.

Thus we find that the 'weak' confinement hypothesis is incorrect.
The QGP can be any colour representation
globally and these are all degenerate. So QGP can be in all or any of 
these colour representations globally.
Colour singlet at large $ T V^{1/3}$ is not favoured.
Now this large ( $\sim 10-20$ fm ) region attained in heavy ion 
collisions at about 200 Mev would condense to $\sim 1 fm$
size colour singlet objects as the system cools.
So as such it should be seen to demonstrate confinement in QCD.
Various coloured representations coexist at
high temperatures globally in QGP. So in this phase the quarks and gluons
in QGP are deconfined. As temperature drops 'local' colour confinement of 
size of $\sim 1 fm$ takes place. 
So the local colour confinement is a result 
of coloured particles in thermodynamic equilibrium at high temperature
undergoing phase transition at a lower temperature. Confinement means 
nothing more than a clear preference for quark and gluons to choose colour 
singlet representation over other colour representations in a size of 
$\sim 1$ fm. In fact we showed that the colour octet and the colour
27-plet move  upto  infinite energies as the temperature drops [7].
This thus is a proof of colour confinement in QCD. In the early universe
when the system was hot the QGP soup was globally in no specific
colour state. As the system cooled and the QCD phase transition took 
place, it created small ( $\sim 1-2$ fm ) size colour singlet objects.
This is what we see at present as colour singlet hadrons.
This is our proof of colour confinement in QCD! 

So as per the QGP scenario above, there was no baryon number in the early 
universe. But today we know that baryon number dominates in the universe.
The fact, that there are no antibaryon present in any significant number 
in the universe now, is referred to as the baryon asymmetry problem. So if 
as per the above  scenario the early universe had no baryon then where did 
the baryon number come from? Below we shall show that the phase transition
itself which demonstrates "strong" colour confinement also leads to 
creation of baryons. We shall see below that this is a baryon number 
violating phase transition.

Fig.1 shows that the colour representation of states: (0,0)-singlet,
(1,1)-octet, (2,2)-27plot are degenerate at high $T V^{1/3}$ 
value. This holds for all flavours: 0-f, 2-f and 3-f cases. 
Now the zero flavour case is the pure $S U (3)_{c} $ gauge 
case where only gluons are present. These correspond to the self conjugate
Young diagrams (0,0),(1,1),(2,2),(3,3) etc and hence are "real".
As the 2-f and 3- f cases are also degenerate with this at 
high $T V^{1/3}$ it implies that all
the states whether colour singlet or not, whether they contain
quarks and antiquarks or not, necessarily correspond to the "real" 
representations. This guarantee of the reality of all relevant 
representatives is significant.
Notice in Fig.1 that at lower values of $T V^{1/3}$ the 0-f case
decouples from the finite quark flavour cases as the singlet for these 
drop down in a phase transition. Note that in the 0-f case the colour 
singlet state is guaranteed to be "real".
But as the 2-f, 3-f cases drop down and
they are decoupled from the 0-f case. Hence these are not 
necessarily restricted by this reality requirement any more.
However the B=0 case is "real" anyway. But
the $q - \bar {q}$ and gluons are all intermixed to give a colour singlet
state. Remember that our quarks are "current" quarks however, now 
intermixed in a sea of gluons. 
As such they are very much looking like the constituent  
quarks at low  $T V^{1/3}$. This picture of constituent quarks arising 
from a sea of quarks-antiquarks and gluons [10] will be sharpened below.

The colour-singlet objects that emerge at small $T V^{1/3}$ in our picture
will consist of a large number of boxes given symbolically as below.

\begin{equation}
\begin{tabular}{|c|c|c|c|c|c|c|c|c|c|c|c|c|c|}
\hline & & & & & & & & & & & & &
\\
\hline & & & & & & & & & & & & &
\\
\hline & & & & & & & & & & & & &
\\
\hline
\end{tabular}
\end{equation}

With the constituent quarks in colour singlet state one requires only

\begin{equation}
\begin{tabular}{|c|c|c|}
\hline
\\
\hline
\\
\hline
\\
\hline
\end{tabular}
\end{equation}

So constituent quarks are basically built from three boxes in a single 
column in $S U (3)_{c} $. 
So when considering single boxes 
( representing constituent quarks ) as above in eq. (2), we are
basically not presenting all the other left out boxes.
So a large number of colour singlet states made up
of current quarks and antiquarks and gluons and consisting of
a large number of columns of three boxes stand
for the sea of the constituent objects in our picture.
     
Before proceeding any further the physical significance of the 
methodology used in our projection technique should be emphasized.
At relativistic energies as the number of particles is not conserved 
hence the concept of canonical distribution should be modified.
The canonical distribution is no longer related to the fixed number of 
particles but rather
to the given irreducible  representation of the symmetry group 
- in our case here the $S U (3) _{c}$  QCD representations [4,2 p. 94]. 
Hence colour confinement in the form of colour singletness in the 
$S U (3)_{c}$ space of QCD is the primary determining factor as 
to what happens and what does not happen here. Particle number is not 
conserved here. What is conserved is the representation.
This has immediate important implications for our results here.
     
The colour singlet state in a single column for B=0 case 
consists of a constituent 
quark and an antiquark state. In the colour space $S U (3)_{c}$
the antiquark is represented by the Young diagram

\begin{equation}
\begin{tabular}{|c|c|c|}
\hline
\\
\hline
\\
\hline
\end{tabular}
\end{equation}

For the antiquark this Young digram also provides the
baryon number B=-1/3. Hence colour
singlet state would represent as per discussion above a constituent quark
and a constituent antiquark to form a colour singlet meson. This
corresponds to a "real" representation.

However the same two box Young diagram also stands for baryon number
B=2/3 for a diquark. This is permitted as per the discussion above.
Hence this diquark state would then along with another constituent quark 
create a baryon number B=1 state. 
Hence because of the dominance of the colour singlet state at low 
$T V^{1/3}$ baryon number violating phase transition can
take place. Now this B=1 Young diagram corresponds to a 
spinor representation. Note that at the same time real representation is 
needed for the B=0 case.

The antiquark with baryon number -1/3 is a fermion while the diquark with
B=2/3 is a boson. Then these two combine with a quark in colour triplet 
state to form states with B=0 and B=1 which are mesons and baryons 
respectively. Hence for B=1 to emerge at low energies along with
the B=0 state it is essential that there should exist a fundamental 
symmetry between the antiquark and the diquark state.
This means that there should be a supersymmetry between a flavour 
$\bar{3}$ multiplet of antiquarks ( $\bar{u}, \bar{d}, \bar{s}$ )
and a flavour $\bar{3}$ multiplet of 
scalar diquarks ( ds-sd, us-su, ud-du ).
These should form the fundamental multiplet of the supersymmetric group
SU(3/3). Indeed such a dynamic supersymmetry between antiquarks and 
diquarks is well known to exist [11]. Starting with the fundamental 
multiplet of the SU(3/3) supersymmetry one then associates the 
pseudoscalar and the vector mesons with
the octet and the decuptet of baryons in a single supermultiplet of the 
supersymmetric algebra SU(6/21). Here the quarks belong to a 6-plet of 
SU(6) and the diquarks to the 21-plet [12,13,14].
This supersymmetry between B=0 and B=1 states has been
well studied [11,12,13,14,15]. It is this supersymmetry which
ensures that in our model the baryon number is generated at small 
$T V^{1/3}$ from a system which was baryonless at higher values. 

As an additional point our model of hadrons 
at low temperatures is reminiscent of the statistical model 
for hadrons [16] and provides a justification for the same. Also as
diquarks seem to be playing a fundamental role in the structure of
hadrons, it may be an indication of deformation of nucleons [17].

On the basis of what has been said a constituent quark would be one
which is represented by a single box. Implicit is the fact that we have 
not shown attached to it a series of a large number of three column boxes
representing the sea of quark- antiquarks and gluon in a colour singlet 
state. So also for antiquark and diquark with a column of two boxes. 
In fact a gluon in the octet state (1,1) may also 
become a constituent gluon by virtue of the fact that it does have a 
minimum in Fig. 1. But that can happen only at higher $T V^{1/3}$ values. 
In fact all the self conjugate representation (2,2), (3,3) may, if allowed 
by the dynamics as represented in Fig 1, become constituent objects. 
However for the colour singlet case, as the supersymmety SU(3/3)
works at the level of a single antiquark and diquark, single column
constituitiveness is all that may arise. That is we cannot associate 
constituent character with colour singlet object with two columns at low
$T V^{1/3}$. With the colour singlet state arising in the 0-flavour case
we have just one complete composite state which cannot be broken into
any sub-constituent parts. 

It is possible that in the early universe there were regions 
free of quarks and antiquarks and were populated entirely by gluons.
Phase transition in that region would have led to colour singlet 
glueballs. As there are no quarks and antiquarks present this 
0-f glueball will not participate in electromagnetic and weak interactions. 
These glueballs may however interact gravitationally 
with each other to create a larger cluster of glueballs.
Also they will not interact with other baryons and mesons 
because as discussed above their "intrinsic" constituent characteristics 
are so very different. 
As these glueballs do not have any quarks, quark exchange force with other 
hadrons is not possible. Also gluon exchange with other hadrons not 
possible as this constituent glueball would by emitting a 
current gluon fluctuate to constituent 
glueball octet state, which can exist only at high $T V^{1/3}$.
In fact as per Fig. 1 ( and [7] ) the 0-f glueball had 
decoupled very strongly from the flavour singlet cases and had a sharper 
and deeper drop. So given an opportunity the glueball colour singlet 
clustering would dominate. 
Ideally speaking the only interaction these glueballs shall partake 
in is the gravitational interaction. So as such these
are ideal candidates for the Dark Matter of cosmology. So QCD itself
( without invoking any exotic extensions of the the Standard Model ) 
is capable of explaining the problem of DM. In fact a couple of years ago 
[18] the the author had looked at the possibility of the octet
representations of the glueball in QCD as a DM candidate.
Now that picture can be further justified as on the basis of the above 
discussion these may act as constituent gluons [18] but at higher 
$T V ^{1/3}$. At lower values it is this new picture which may dominate. 
Also our picture can explain as to
why it is so hard to produce glueballs in the laboratory. This has to do 
with the fact that in the laboratory e.g. in the Heavy Ion Collisions
it is not possible to create any substantial quark-antiquark free region.
     
Hybrids are made up of ( $ q \bar {q} g $ ) configuration.
Exotics like these have not been created in the laboratory so far.
To form this exotic, say ($q \bar {q}$) subsystem would have to be in 
colour octet state which along with the gluon octet 
would produce a colour singlet state.
But both these representations are suppressed at low $T V^{1/3}$ in our 
picture. Hence it would be hard to produce it in the laboratory.
But in Fig.1 note that in both these cases there is indeed a minimum at 
higher $T V^{1/3}$ values. Thus it may be possible
to produce hybrids at higher temperatures and also as these may appear as 
"constituent" particles there. We have already predicted
existence of hybrids as a possible signature of QGP at higher
temperatures [19]. The reluctance of other exotics to show up in the
laboratory can easily be explained in our model.
So diquonia $(q \bar {q}) (q \bar {q})$ to exist as a single composite 
would require each quark-antiquark pair for example to be in an octet 
state. But such representations are suppressed in our picture and these 
hybrids would not be produced in the laboratory. 
     
The above was true for the QGP in the early universe scenarios. What
does this new perspective prove for the experiments conducted in the
laboratory. As per our picture if we start with two heavy ions each with
their baryon numbers A, A' then when they collide and form QGP then in 
this phase the baryon numbers will disappear. 
Thus in the QGP here too baryon
numbers will disappear. In this phase the system will globally exist in 
any arbitrary colour representations. The QGP would be confined in a finite
size of $\sim 10-20$ fm for a short period 
due to the dynamics of the collision. As
the system cools baryons would be regenerated and due to (global) baryon
number conservation and due to the forward and the backward current 
conservation at sufficiently higher energies the same baryons would 
continue to move out. This will happen as the other
colour singlet hadrons with overall B=0 would be leaking out all over. 
And this is exactly what is seen in heavy ion collisions! Hence our 
picture is vindicated.

Note that as per our model indeed QGP has already been formed in the
laboratory. Baryonlessness of the same is a signal for the
formation of QGP. Our model can be interpreted in the framework
of the "cold plasma" suggested sometime ago [20].
We feel that the 2-10 GeV/nucleon collisions in the centre of mass as
observed by the PLASTIC ball detector [21 p. 108,110] has already 
observed the QGP in the laboratory.
At higher energies the so called "nuclear transparency" is a post
QGP phase wherein as per the conservation of the baryonic currents
one sees the baryons continuing to move forward and backward.
There is no transparency really. In the QGP phase the whole system had 
lost all baryon number and it got generated later on in the cooling phase.
Also the so called 'colour transparency' observed in
p-p and pA collisions should be naturally explained in terms of our 
picture here. The regions of high multiplicity fluctuations in 
hadron-hadron collision are regions where QGP with B=0 are formed.
The proton number continues through, so to say, as it melts its way 
through the hadrons. Detailed calculations shall have to be done to 
consolidate this new understanding of these effects.
In this paper the aim has been to show the power of the methodology 
adopted to explain and correlate several puzzling effects and phenomena in 
hadronic physics within a consistent and coherent picture arising from
a proper understanding of confinement in QCD.

\newpage

{\bf Appendix}

\vspace{.1in}

The colour projection technique uses
the orthogonality relation for the associated characters 
$ \chi_{(p,q)} $ of the (p,q) multiplet of the group $ SU(3)_c $
with the measure function $ \zeta(\phi,\psi) $ is

\begin{equation}
\int_{SU(3)_{c}}^{} d \phi \, d \psi \, \zeta \left( \phi,\psi \right) 
\chi^{\star}_{(p,q)}\left( \phi,\psi \right) 
\chi_{(p',q')} \left( \phi,\psi \right) = \delta_{pp'} \delta_{qq'}
\end{equation}

The generating function
$ Z^{G} $, the canonical partition function $ Z_{(p,q)} $ are

\begin{equation}
Z^{G}(T,V,\phi,\psi) = \sum_{p,q} \frac{ Z_{(p,q)} }{ d(p,q) }
\chi_{(p,q)} ( \phi,\psi )  
\end{equation}

\begin{equation}
Z_{(p,q)}=tr_{(p,q)} \left[ 
\exp{ \left( -\beta \hat{H}_{0} \right) } \right]
\end{equation}

The many-particle-states which belong to a given multiplet (p,q)
are used in the statistical trace with the 
free hamiltonian $\hat{H}_0$, d(p,q) is
its dimensionality and $\beta$ is the inverse of the temperature T.
The projected partition function $ Z_{(p,q)} $
is obtained by the orthogonality relation for the 
characters. Hence the projected partition function for any representation
(p,q) is
\begin{equation}
Z_{(p,q)}= d(p,q)
\int_{SU(3)_{c}} d\phi \, d\psi \, \zeta \left( \phi,\psi \right) 
\chi^{\star}_{(p,q)}(\phi,\psi)
Z^G \left( T,V,\phi,\psi \right)
\end{equation}

The characters of the different representations are given below: \\
\begin{eqnarray}
 \chi_{(1,0)} & = & exp{(2i\psi/3)} + 
                      2exp{(-i\psi/3)}cos(\phi/2) \\
  \chi_{(0,1)} & = & \chi_{(1,0)}^{\star} \\
  \chi_{(1,1)} & = & 2 + 2 \left[ cos\phi 
    +  cos\left( \phi/2 + \psi \right) 
    +  cos\left(-\phi/2 + \psi \right) \right] \\
  \chi_{(2,2)} & = & 2 + 2 \left[ cos \phi  
    +  cos (3\phi/2) cos (\phi/2) \right] + \nonumber \\
               &   & 2 \left( 1+2cos\phi \right) 
       \{ cos\left( \phi/2 + \psi \right) 
    +         cos\left(-\phi/2 + \psi \right) 
            + cos 2 \psi + \left( 1/2 \right) cos \phi 
       \} 
\end{eqnarray}
The expressions of the generating function used above is 
\begin{equation}
 Z^G(T,V,\phi,\psi) 
= tr\left[exp(-\beta \hat{H}_{0}
         + i\phi \hat{I}_z + i\psi \hat{Y}) \right] 
\end{equation}
where $\hat{I}_{z}$ and $\hat{Y}$ are the diagonal generators of the 
maximal abelian subgroup of $SU(3)_c$. 
Our plasma consists of light spin 1/2 quarks and antiquarks in the 
triplet and antitriplet representations (0,1) and (1,0) respectively, and 
massless spin one gluons in the octet representation (1,1). 
Note that the non-interacting 
hamiltonian $\hat{H}_0$ is diagonal in the occupation-number
representation.
In the same representation one can write the charge 
operators $\hat{I}_z$ and $\hat{Y} $ as linear 
combinations of particle-number operators. Hence $Z^G$
can be easily calculated
in the occupation-number representation. With an imaginary 
`chemical potential' this is just like a grand canonical
partition function for free fermions and bosons. One obtains
\begin{eqnarray}
Z^G_{quark}=\prod_{q=l,m,n} \prod_{k} 
   \left[ 1 + exp{( -\beta\epsilon_k - i\alpha_q) } \right] 
   \left[ 1 + exp{( -\beta\epsilon_k + i\alpha_q) } \right] \\
Z^G_{glue}=\prod_{g=\mu,\nu,\rho,\sigma} \prod_{k} 
   \left[ 1 - exp{(-\beta\epsilon_k+i\alpha_g)}  \right]^{-1}
   \left[ 1 - exp{(-\beta\epsilon_k-i\alpha_g)}  \right]^{-1}
\end{eqnarray}

The single-particle energies are given as 
$\epsilon_k$. For (1,0),(0,1) and (1,1) multiplets, 
the eigenvalues of 
$ \hat{I}_z $ and $ \hat{Y} $ give different angles as:
\begin{eqnarray}
\alpha_l=(1/2) \phi + (1/3) \psi, \alpha_m=(-1/2) \phi + (1/3) \psi,
  \alpha_n=(-2/3) \psi \\
\alpha_{\mu}=\alpha_l-\alpha_m, \alpha_{\nu}=\alpha_m-\alpha_n,
  \alpha_{\rho}=\alpha_l-\alpha_n, \alpha_{\sigma}=0
\end{eqnarray}

We neglect the masses of the light quarks. 
At large volume the 
spectrum of single particle becomes  a quasi-continuous one and
$\Sigma ... \rightarrow V/({2\pi})^{3} \int d^{3}p .. $. 
\begin{equation}
Z^G (T,V,\phi,\psi) = 
  Z_{quark}^G (T,V,\phi,\psi) Z_{glue}^G (T,V,\phi,\psi) 
\end{equation}

This then enables us to obtain the partition function for any 
representation ie. $Z_{(p,q)}$. 
One may thus obtain any thermodynamical quantity
of interest for a particular representation. For example
the energy 
\begin{equation}
 E_{(p,q)} = T^2 \frac{\partial}{\partial T} ln Z_{ (p,q) } .
\end{equation}

Here we project out different representations like singlet (0,0),
octet (1,1), 27-plet (2,2) etc. in QGP. 
Let us look at singlet, octet, 27-plet etc. projection. 
Take $\mu = 0$ for 0, 2 and 3 flavours. Next plot in Figure 1:
\begin{equation}
D_{(p,q)}^{eff} = E_{(p,q)}/E_0 = 1 + E_{(p,q)}^{corr}/E_0
\end{equation}

\newpage

{\bf References} 

\vspace{.2in}

1. A. Smilga, {\it "Lectures on Quantum Chromodynamics"}, 
   World Scientific, Singapore,2001

2. B. M\"{u}ller, {\it "The Physics of Quark-Gluon Plasma"}, 
   (Lecture Notes in Physics, Vol. 225), Springer Verlag Berlin 1985

3. N.K. Glendenning,
   {\it "Compact Stars"}, Springer-Verlag, New York 1997

4. K. Redlich and L. Turko, 
   {\it Z. Phys.} {\bf C 5} (1980) 201;
   L.Turko, {\it Phys. Lett.}, {\bf B 104} (1981) 153

5. M.I. Gorenstein, S.I. Lipskikh and G.M. Zinovjev,
   {\it Z. Phys.}, {\bf C 22} (1984) 189

6. H.-Th. Elze, W. Greiner and J. Rafelski,
   {\it Phys.Letts.}, {\bf B 124} (1983) 515

7. Afsar Abbas, Lina Paria and Samar Abbas, {\it Eur. Phys. J.}, 
   {\bf C 14 } (2000) 695

8. H.-Th. Elze, W. Greiner and J. Rafelski, 
   {\it Z. Phys.}, {\bf C 24} (1984) 361

9. G. Auberson, L. Epele, G. Mahoux and F.R.A. Sima\~{o},
   {\it J. Math. Phys.}, {\bf 27} (1986) 1658

10. G. Altarelli, N. Cabibbo, L. Maiani and R. Petronzio,
   {\it Nucl. Phys.}, {\bf B 69} (1974) 531

11. M. Anselmino, E. Predazzi, S. Ekelin, S. Fredriksson and 
   D.B. Lichtenberg, {\it Rev. Mod. Phys.}, {\bf 65 } (1993) 1199

12. H. Miyazawa, {\it Phys. Rev.}, {\bf 170 } (1968) 1586

13. S. Catto and F. Guersey, {\it Nuovo Cim.}, {\bf A 86} (1985) 201; 
   {\it Nuovo Cim.}, {\bf A 99} (1988) 685

14. R.C. Hwa and C.S.Lam, {\it Phys. Rev.}, {\bf D 12} (1975) 3730;
    {\it Phys. Rev.}, {\bf D13} (1976) 2096

15. D.B. Lichtenberg, {\it J.Phys.}, {\bf G 16} (1990) 1599

16. A. Bhattacharya and S.N. Banerjee,
    {\it Prog. Theor. Phys}, {\bf 81} (1989) 555

17. Afsar Abbas, {\it J.Phys.}, {\ G 18} (1992) L89

18. Afsar Abbas, {\it "A new candidate for the Dark Matter"},
    hep-ph/9504430

19. Afsar Abbas and Lina Paria, {\it J.Phys.}, {\bf G 23} (1997) 791

20. H. Stoecker and W. Greiner, {\it Phys. Rep.}, {\bf 137} (1986) 277

21. K. Heyde, {\it "From Nucleons to the Atomic Nucleus"}, 
    Springer Verlag, Berlin, 1997

\newpage

\begin{figure}
\caption{
$ D_{eff} $ for the colour representations: singlet and octet as a 
function $ TV^{1/3} / \hbar c $  
for different number of flavours: 0 (pure gauge case), 2 and 3. }
\epsfclipon
\epsfxsize=0.99\textwidth
\epsfbox{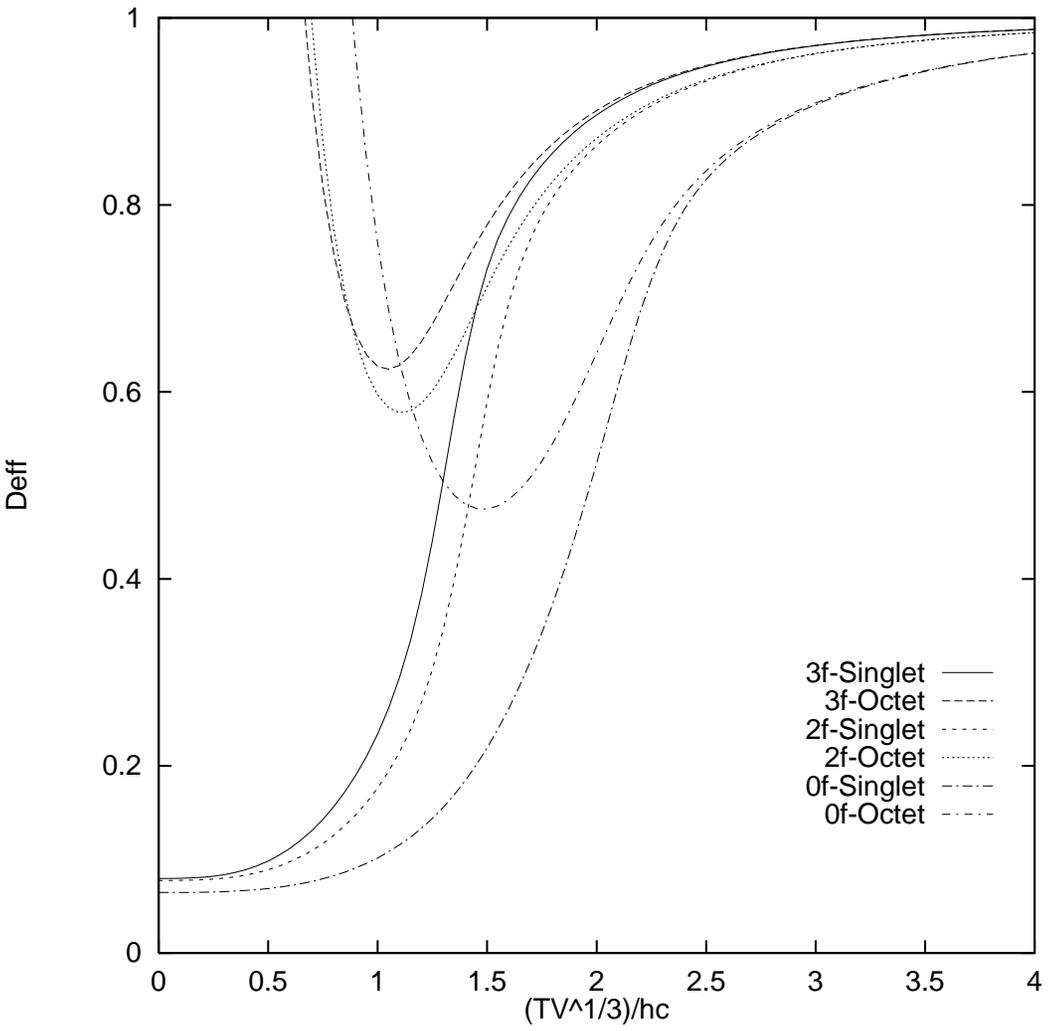}
\end{figure}

\end{document}